\documentclass{emulateapj}

\slugcomment{Accepted for publication in  {\em The Astrophysical Journal Letters}.}

\shortauthors{Shuping et al.}
\shorttitle{Silicate Emission from Protostellar Disks in Orion}
\newcommand{\thetaonec}{$\theta^{1}$Ori C}
\begin{document}

%TITLES-------------------------------
\title{Silicate Emission Profiles from Low-Mass Protostellar Disks in the Orion Nebula:  Evidence for Growth and Thermal Processing of Grains}

\author{R. Y. Shuping\altaffilmark{1}, 
Marc Kassis\altaffilmark{2},
Mark Morris\altaffilmark{3},
 Nathan Smith\altaffilmark{4}, \& 
 John Bally\altaffilmark{4}
 }

\altaffiltext{1}{Universities Space Research Assoc.,
 NASA Ames Research Center,
 MS 211-3,
 Moffett Field, CA   90035 \\ {\it email}: {\tt rshuping@sofia.usra.edu}}
 
\altaffiltext{2}{W. M. Keck Observatory, 
65-1120 Mamalahoa Hwy., 
Kamuela, HI 96743 \\
{\it email}:  {\tt mkassis@keck.hawaii.edu}}

\altaffiltext{3}{Div. of Astronomy \& Astrophysics, 
Univ. of California, 
Los Angeles, CA 90095 \\
{\it email}:  {\tt morris@astro.ucla.edu}}

\altaffiltext{4}{Center for Astrophysics \& Space Astronomy, 
Univ. of Colorado, 
389 UCB, 
Boulder, CO 80309 \\
{\it email}:  {\tt nathans@casa.colorado.edu} \\
{\it email}:  {\tt bally@origins.colorado.edu}}

%------------------------------------

%ABSTRACT------------------------------
\begin{abstract}

We present 8--13~\micron\ low resolution spectra ($R\approx100$) of 8 low-mass protostellar objects (``proplyds'') in the Orion Nebula using the Long Wavelength Spectrometer (LWS) at the W.~M. Keck Observatory.  All but one of the sources in our sample show strong circumstellar silicate emission, with profiles that are qualitatively similar to those seen in some T Tauri and  Herbig Ae/Be stars.  The silicate profile in all cases is  significantly flattened compared to the profile for typical interstellar dust, suggesting that the dominant emitting grains are significantly larger than those found in the interstellar medium.  The 11.3-to-9.8 \micron\ flux ratio---often used as an indicator of grain growth---is in the 0.8 to 1.0 range for all of our targets, indicating that the typical grain size is around a few microns in the surface layers of the attendant circumstellar disk for each object.  Furthermore, the silicate profiles show some evidence of crystalline features, as seen in other young stellar objects.  The results of our analysis show that the grains in the photoevaporating protostellar disks of Orion have undergone significant growth and perhaps some annealing, suggesting that grain evolution for these objects is not qualitatively different from other young stellar objects.  

\end{abstract}
%---------------------------------------


%MAIN BODY------------------------------

\section{Introduction}
\label{sect:Intro}
 
Disks surrounding young stellar objects are critical to our understanding of star formation.  Not only do circumstellar disks provide a mechanism for angular momentum transport and accretion onto the central star, but they also serve as the birthplace of planets.  Many observed disks exhibit prominent emission at 9--12~\micron\ originating from the Si--O stretching mode of silicates present in the dust grains.  These silicates are primarily amorphous, but crystalline minerals (e.g. forsterite and olivine) are also observed.  This emission is thought to form in the warm, optically thin atmosphere of the inner accretion disk heated by the central protostar~\citep{Chiang:1997,Natta+:2000}.  The relative strength and shape of the silicate profile is a function of the grain composition and grain size~\citep{Henning+95,Bouwman+01}.  Recent studies of Herbig Ae/Be (HAeBe) stars, T Tauri stars, and brown dwarfs suggest that the grains in the surface layers of the disks around these objects have undergone significant thermal processing and have grown well beyond typical interstellar medium sizes~\citep{Przygodda+03,vanBoeckel+03,Kessler-Silacci+05, Apai:2005}---suggesting that the initial phases of planet formation could be underway in these objects.

\begin{figure*}
%\begin{center}
\epsscale{1.1}
%\plotone{trap_IR-HST.eps}
%\includegraphics[0in,0in][6.5in,6.5in]{f1_rev1.jpg}
\plotone{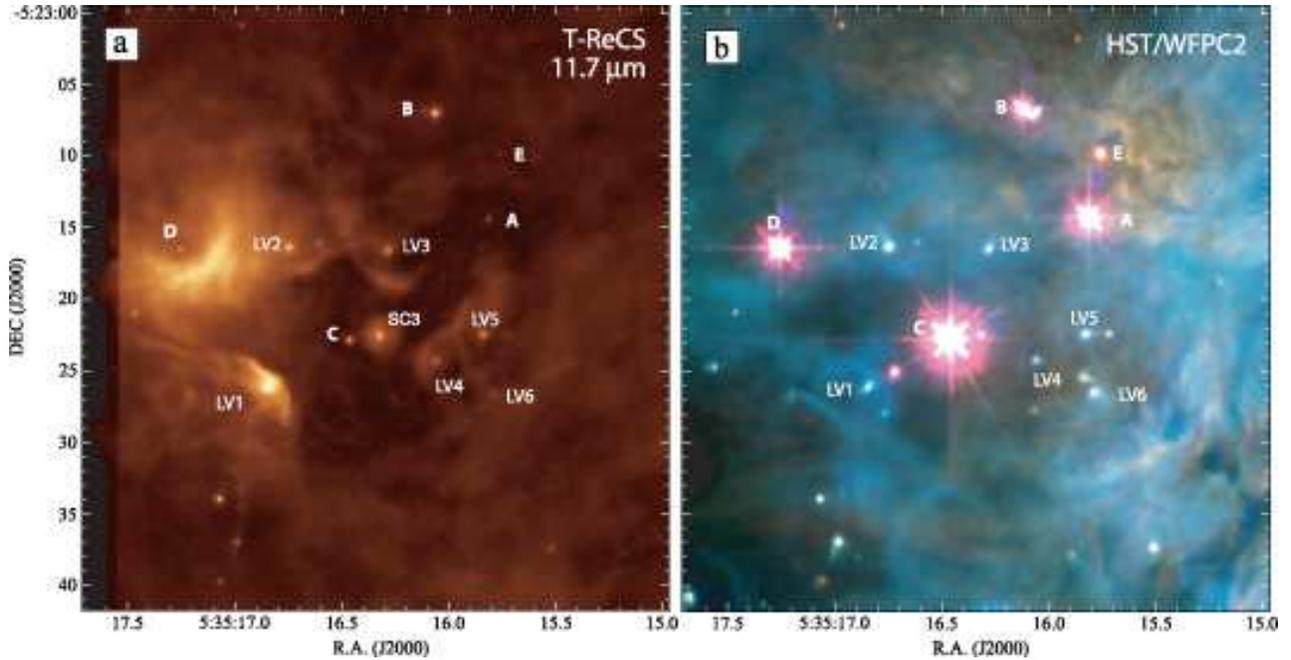}
\caption{Detail of area immediately surrounding the Trapezium stars in Orion. (a) False-color  11.7~\micron\ image taken with the T-ReCS instrument at the Gemini South Observatory~\citep{Smith:2005}.  (b) HST/WFPC2 composite image (Blue = [O III]; Green = H$\alpha$; Red = [N II]).  The trapezium stars ($\theta^{1}$ Ori) are marked A -- D along with the proplyds observed for this work; SC3 and LV 1 -- 6.  HST3 is $\approx$30\arcsec\ south of \thetaonec, just off this field of view.  Figure from \citet{Smith:2005}.}
\label{fig:trap_IR-HST}
%\end{center}
\end{figure*}

The {\it Hubble Space Telescope} has produced dramatic images of protoplanetary disks (``proplyds'') surrounding young stars in the Orion Nebula~\citep{ODell+93,Bally+98,Bally+00} (Figure~\ref{fig:trap_IR-HST}, panel b).  The intense ultraviolet (UV) radiation field of the high-mass Trapezium stars heats the disk surfaces to a few thousand degrees, drives mass-loss, and produces bright ionization fronts~\citep{Johnstone+98}.  Proplyds are thought to contain young ($< 10^6$~year old) low-mass stars which are part of the Orion Nebula Cluster~\citep{Hillenbrand97}.  

%Trap Figure -----------------------------------
%---------------------------------------------

The photoevaporation of these disks places strong temporal constraints on planet formation mechanisms in irradiated environments near O-type stars; which is important since a significant fraction of stars forms in clusters with high-mass stars~\citep{Blaauw91}.
Furthermore, recent meteoritic studies suggest that primitive solar system materials were subjected to intense UV irradiation~\citep{Lyons:2005} and that the whole protoplanetary disk may have been polluted by the ejecta from a nearby supernova explosion~\citep{Tachibana:2003,Hester:2004}---both suggesting that our own solar system may have formed in a region similar to Orion. So it is clearly important to gain an understanding of how efficient planet formation is for low-mass stars forming in clusters with high-mass members, both to predict the frequency of planetary systems and to better understand the evolution of our own solar system.  

  Recent studies show that grain growth has occurred in the giant silhouette disk of proplyd 114-426~\citep{Throop+01,Shuping+03}, suggesting that the processes leading to planetesimal formation are underway.  Recent numerical modeling has shown that grain growth to centimeter sizes and beyond can occur in the mid-plane of these photoevaporating disks~\citep{ThroopThesis:2001}.  Furthermore, planetesimal formation via gravitational instability may in fact be {\it enhanced} by the photoevaporation of the disk surface layers~\citep{Throop+Bally05}---leading to the somewhat paradoxical conclusion that planet formation around low-mass stars may in fact be {\em more efficient} in the harsh environments of high mass clusters.  In this letter, we present initial results from a 8--13~\micron\ spectroscopic survey of the Orion proplyds.  Most of the proplyds in our survey exhibit strong silicate emission profiles clearly showing grain growth beyond typical ISM sizes in the surface layers of these externally illuminated, photoablating circumstellar disks.

\section{Observations \& Analysis}
\label{sect:Obs}

\begin{figure*}
%\begin{center}
\epsscale{1.1}
%\plotone{sil_profiles_rev1.eps}
\plotone{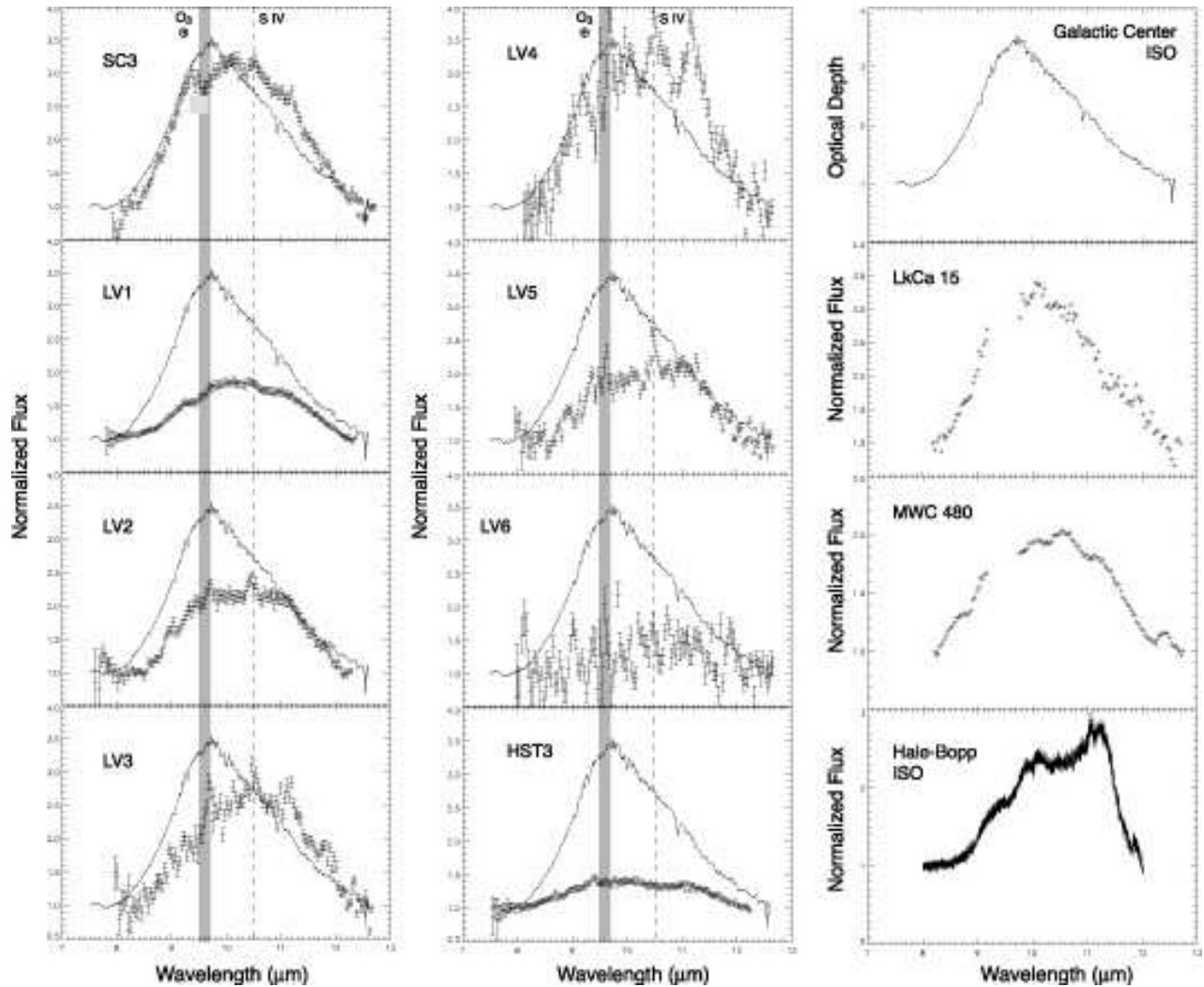}
\caption{Silicate emission profiles for proplyds in Orion and some comparison spectra:  The left 2 columns show continuum subtracted and normalized Keck/LWS spectra from 8 to 13~\micron\ for the proplyds in our sample (crosses with 1$\sigma$ error bars; see text for description of normalization procedure) with the optical depth spectrum toward the Galactic Center overlaid for comparison (line; see below).  The gray strip indicates the location of the deep atmospheric ozone band at 9.6~\micron; the dashed line indicates the location of [\ion{S}{4}] background nebular emission at 10.5~\micron.  In theory, the nebular [\ion{S}{4}] should be eliminated by sky subtraction; but in practice, spatial and spectral variability in the line can cause large residuals in the final spectra.  The right hand column shows some comparison profiles.  The Galactic Center and Hale-Bopp spectra were obtained from the ISO Data Archive; while spectra for the T Tauri star LkCa~15 and the HAeBe star MWC~480 are from \citet{Kessler-Silacci+05}.  The silicate band is observed in absorption towards the Galactic Center---we present the optical depth here to facilitate comparison to the emission profiles.}
\label{fig:sil_profiles}
%\end{center}
\end{figure*}

As part of our ongoing studies to understand the grain composition and size distribution in the proplyd disks, we have obtained 8--13~\micron\ low resolution spectra (R$\approx$100) of 8 targets  in Orion using the Long Wavelength Spectrometer (LWS) at the W. M. Keck Observatory~\citep{Jones+Puetter93} on UT 21 and 24 December 2004.  All observations were made using standard chop-nod techniques. Telluric absorption features were removed by dividing our object data by a standard star and multiplying by a calibrated spectral template~\citep{Cohen:1999}.  The wavelength solution was determined using sky lines throughout the 10~\micron\ window.  Continuum subtracted and normalized spectra for the 8 targets are shown in Figure~\ref{fig:sil_profiles}. The normalized flux, $S_{\nu}$, for each source was determined using the following formula:
\begin{equation}
S_{\nu} = 1 + \frac{F_{\nu} - F_{c,\nu}}{\langle F_{c,\nu} \rangle}
\end{equation}
where $F_{\nu}$ is the observed flux, $F_{c,\nu}$ is the linear continuum fit to the observed fluxes at 8 and 12.5~\micron, and $\langle F_{c,\nu} \rangle$ is the frequency average of the fitted continuum.  This procedure effectively removes any slope from the silicate profile imposed by the underlying continuum, facilitating comparison to other observed and model profiles reported in the literature.  

All sources in our sample except LV6 (discussed below) show prominent silicate emission profiles that are qualitatively similar to those observed for some T Tauri and HAeBe stars. For comparison, spectra of the Galactic Center sightline~\citep[e.g.][]{Kemper:2004} and comet Hale-Bopp~\citep[e.g.][]{Wooden:1999}---both obtained from the ISO Data Archive\footnote{
{\tt http://isowww.estec.esa.nl/}}
---along with the T Tauri star LkCa~15 and the HAeBe star MWC~480~\citep{Kessler-Silacci+05} are shown in Figure~\ref{fig:sil_profiles}.   It is important to note, however, that profile strength and shape vary widely among HAeBe and T Tauri stars~\citep[e.g.]{Bouwman+01,Przygodda+03}; we have deliberately picked examples that appear similar to the proplyds observed in our sample.  Since the sightline passes through $A_V \sim 30$ magnitudes of primarily diffuse interstellar material, the silicate spectrum towards the Galactic Center is often used as a proxy for typical diffuse interstellar dust composed primarily of amorphous silicates.  The profile for Hale-Bopp, on the other hand, shows strong crystalline silicate features, especially forsterite at 11.3~\micron, thought to be produced by thermal annealing of the grains~\citep{Fabian:2000,Davoisne:2006}.

%Silicate Profiles -----------------------------------
%---------------------------------------------

The absorption coefficient for amorphous olivine with typical ISM grain sizes (radius $a\sim0.1$~\micron) is strongly peaked at 9.8~\micron\ (as seen in the Galactic Center profile); but the coefficient flattens between 9.5 and 12~\micron\ for larger grain sizes with the same silicate composition ($a\sim2.0$~\micron)~\citep{Bouwman+01}.  Detailed optical models of silicate grains show that the 10~\micron\ feature decreases significantly in peak strength and flattens out as the grain size increases from 0.1~\micron\ to 6.0~\micron\ (for constant mass); the feature practically disappears for grain sizes $> 8.0$~\micron~\citep{Kessler-Silacci+05}.  Hence, the ratio of 11.3 to 9.8~\micron\ flux is often cited as a strong indicator of grain growth~\citep{Przygodda+03,Kessler-Silacci+05}, where values of 0.5 indicate grains with $a \sim 0.1$~\micron, and values of around 1.0 imply $3.0<a<8.0$~\micron .  \footnote{
Recent studies indicate that the silicate profile is also sensitive to the details of grain porosity and shape; this leads to an {\em underestimate} of grain size as determined from the 11.3-to-9.8 \micron\ flux ratio~\citep{Min:2006,Voshchinnikov:2006}.}
The sources in our sample have $F(11.3)/F(9.8)$ ratios in the 0.75 to~1.0 range indicating grain sizes of at least a few microns.

There is significant structure in the profiles as well, suggesting the presence of crystalline silicates.  Most of these sources have smaller peaks at 9.2 and 11.3~\micron\ indicative of crystalline enstatite and forsterite, respectively. The peak at 10.5~\micron\ seen in some sources might also be due to enstatite, but this cannot be confirmed due to the possibility of [\ion{S}{4}] emission at the same wavelength from the proplyd or nebular background. For LV4 and 5, the 11.3~\micron\ peak is starting to dominate the profile, creating a shape similar to that seen in comet Hale-Bopp.  The presence of crystalline silicates indicates that the grains have undergone some thermal processing---usually attributed to annealing of the grains in the inner region of the circumstellar disk~\citep{Bouwman+01}.  Since the proplyd disks are heated to $\sim1000$~K at the disk surface by the UV radiation field of \thetaonec, it is possible that the grains are undergoing thermal annealing across the entire side of the disk facing \thetaonec.

\section{Discussion}
\label{sect:Discussion}

The infrared emission from dust in the proplyds is driven largely by the ultraviolet radiation field from \thetaonec .  For a low-mass protostar near the Trapezium, the disk radius at which heating by the central star is equivalent to heating by the nebular environment is just a few AU for a wide range of inclination angles~\citep{Robberto:2002}.  The total flux density in the silicate lines we observe for the proplyds is roughly similar to that seen in T Tauri stars in Taurus-Aurigae~\citep{Przygodda+03,Kessler-Silacci+05}, despite the fact that Orion is nearly 3.3 times more distant than Taurus-Aurigae.  Assuming that the dust surface density is similar, this implies that the silicate emitting region is roughly 10 times larger for the proplyds.  Since the proplyds are thought to contain low-mass stars---like T Tauri objects---this clearly indicates that UV heating by \thetaonec\ is important.  Since the disks are optically thick at 10~\micron, thermal and silicate emission can arise across the entire surface of the disk and will be strongest for the side that faces \thetaonec ; i.e. proplyds which lie {\em behind} \thetaonec\ on our line of sight will be brightest.  LV6 is the lone target we observed that showed little or no evidence of silicate emission. In addition, it is much less bright at 10 and 20~\micron\ than the other targets (see Fig.~\ref{fig:trap_IR-HST}), suggesting that it is the only one of them in front of \thetaonec\ on the line of sight.

The results of our 8--13~\micron\ spectral analysis clearly indicate that the grains in proplyd disks have undergone significant growth (usually attributed to agglomeration) and perhaps some thermal annealing.  It appears that grain evolution in the proplyd disks is not qualitatively different from that in other protostellar disks, despite ongoing photoablation by the UV radiation from \thetaonec.

In general, the grains in circumstellar disks around young stars grow rapidly to centimeter sizes via agglomeration in turbulent eddies, at which point they begin to settle to the disk mid-plane~\citep{Mizuno:1988,Weidenschilling:1997}.  In fact, the coagulation of small grains into centimeter-sized particles and larger is so efficient that unless fragmentation of larger grains is invoked nearly all micron-sized grains are swept up in $<<10^5$~years---which clearly contradicts observed mid-IR SEDs for T Tauri and HAeBe stars~\citep{Weidenschilling:1997,Dullemond:2005}.  Despite fragmentation, a significant fraction of grains progress beyond the 1~cm mark, suggesting that larger bodies can continue to grow at least to sizes of $\sim 1$ meter, where planetesimal formation process is impeded again due to fast radial migration of meter-sized objects toward the host protostar~\citep[e.g][]{Weidenschilling:1997}.  

Grains in the protostellar disks of Orion can also be entrained in the photoevaporative flow driven by the UV radiation from \thetaonec.  Grains at the surface of the disk are acted on by drag forces associated with the evaporating gas and gravity due to the central star and disk:  If the drag forces are great enough, the grain is entrained in the flow.  Emission at 11~\micron\ from arcs around proplyds near \thetaonec\ and HH jets from proplyds imply that small grains are in fact being entrained in these
outflows~\citep[, see Fig.~\ref{fig:trap_IR-HST}]{Smith:2005}.  Whether any particular grain at the disk surface is entrained or not depends on both the particle size  and the disk radius at which it lies: at a radius of 1~AU, grains up to a micron in size can be lost; but at 100~AU, grains up to 50~\micron\ in size are entrained~\citep{ThroopThesis:2001}.  Entrainment, however, should be slow compared to the growth processes for the inner disk regions ($R < 40$~AU): Even though small grains are rapidly depleted from the outer regions of the disk, \citet{ThroopThesis:2001} found that centimeter to meter sized bodies can still form in the disk mid-plane out to 100~AU within $10^5$~years---well within the lifetime of the Orion Nebula Cluster~\citep{Hillenbrand97}.  The growth of grains  up to a few microns (lower limit) in the surface layers of the disk inferred from our silicate profiles indicates that entrainment is indeed much less efficient than coagulation in removing small grains from the inner disk.  Furthermore, since the proplyds are roughly $5\times10^5$~years old, the existence of micron-sized grains at the surface supports the conclusion that fragmentation processes are required to explain the observed grain size distributions in protostellar disks.

Our observations clearly show that grains larger than ISM sizes can grow and survive in the proplyd disks.  Whether planets are being formed is still not known; but grain growth in the surface layers {\it is} a sign of particle evolution leading to planetesimal mid-plane formation in the mid-plane~\citep[e.g.][]{Weidenschilling:1997}.    Further observations at longer wavelengths---where we can peer into the mid-plane of the disk~\citep[e.g.][]{Rodmann:2006}---are required to determine if these planet formation processes are underway.

%SAMPLE DELUXETABLE---------------------------------------
%\begin{deluxetable}{lccccccl}
%\tablewidth{0pt}
%\tablecaption{Target \& Sightline Summary\label{tab_targets}}
%\tablehead{
%\colhead{Target}  &  \colhead{Common}  &  \colhead{Stellar} &
%\colhead{$E$(B-V)} & \colhead{$R_V$}  & \colhead{$A_V$} & \colhead{$D$
%(pc)\tablenotemark{a}} &   \colhead{Notes} \\
%}
%\startdata					
%TABLE DATA GOES HERE
%\enddata
%\tablenotetext{a}{To be used with \tablenotemark{a}}
%\tablecomments{Comments go here.}
%\end{deluxetable}
%-------------------------------------------

%\onecolumn
%\small

\acknowledgements

The authors would like to thank Dr. Jacqueline Kessler-Silacci for graciously providing her data for the comparison spectra.  This work was supported by the Colorado Center for Astrobiology and the UCLA Center for Astrobiology, both supported by the NASA Astrobiology Institute. The data presented herein were obtained at the W.M. Keck Observatory, which is operated as a scientific partnership among the California Institute of Technology, the University of California and the National Aeronautics and Space Administration. The Observatory was made possible by the generous financial support of the W.M. Keck Foundation.  Some of the observations for this research were provided by the W. M. Keck Observatory using DirectorÕs discretionary time, also known as ``Team Keck.''  The authors wish to recognize and acknowledge the very significant cultural role and reverence that the summit of Mauna Kea has always had within the indigenous Hawaiian community.  We are most fortunate to have the opportunity to conduct observations from this mountain.

\end{document}